\pdfoutput=1
\documentclass[11pt]{article}
\usepackage[pdftex]{graphicx,color} 
\usepackage{jheppub}
\usepackage{amsmath}
\usepackage{amssymb}
\usepackage{comment}
\usepackage{multirow}
\usepackage{mathtools}

\usepackage{epsfig,euscript,xspace, color}
\usepackage{mathtools}
\usepackage{slashed} % For Dirac operators 

\def\bea{\begin{eqnarray}}
\def\eea{\end{eqnarray}}
\def\be{\begin{equation}}
\def\ee{\end{equation}}
\newcommand{\nn}{\nonumber\\}

\newcommand{\Log}[1]{\text{ ln}\left(#1\right)}

\newcommand{\bz}{\bar{z}}

\definecolor{lightblue}{rgb}{.1,.4,.5}
\definecolor{brown1}{rgb}{.64,.43,.138}

\def\beq{\begin{equation}}
\def\eeq{\end{equation}}

\def\to{\rightarrow}

\def\DD{\Delta}

\def \f{\phi}

\usepackage{dsfont}

\def\bsub{\begin{subequations}}
	\def\esub{\end{subequations}}

\newcommand{\expec}[1]{\langle #1 \rangle}
\newcommand{\NO}[1]{{:\!#1\!:}}

\def\zb{\bar{z}}

\def\half{\frac{1}{2}}

\def\DD{\Delta}
\def\Oo{\mathcal{O}}

\def\l{\ell}

\def\unit{\mathds{1}} % needs dsfont package

\title{\boldmath Analytic Bootstrap for Logarithmic  CFT}

\author[a, b]{Pinaki Banerjee}  
\author[c]{and Parijat Dey}

\affiliation[^a]{International Centre for Theoretical Sciences\\
Tata Institute of Fundamental Research \\
Shivakote, Bengaluru 560 089, India}

\affiliation[^b]
{Indian Institute of Technology Kanpur\\ 
Kalyanpur, Kanpur 208016, India}

\affiliation[^{c}]{Department of Physics and Astronomy, Uppsala University \\
Box 516, SE-751 20 Uppsala, Sweden\\}

\emailAdd{pinaki.banerjee@icts.res.in, pinakib@iitk.ac.in}
\emailAdd{parijat.dey@physics.uu.se}

\abstract{We study logarithmic conformal field theory (LogCFT) in four dimensions 
	using conformal bootstrap techniques in the large spin limit. We focus on the constraints imposed by conformal symmetry on the four point function of certain logarithmic scalar operators and compute the leading correction to the anomalous dimension of double trace operators in the large spin limit. There exist certain holographic duals to such LogCFTs, which involve higher derivative equations of motion. The anomalous dimension is related to the binding energy of a state where two scalars rotate around each other with a large angular momentum.  We compute this energy shift and compare it to the anomalous dimension of the large spin double trace operators due to stress tensor exchange in the LogCFT. Our result shows that the cluster decomposition principle is satisfied for LogCFTs as long as the dimensions of the operators are positive.
}

\begin{document}

	\maketitle
	\flushbottom

  \section{Introduction}     
 
  Conformal field theories (CFTs) are quite essential in different branches of physics - particularly in statistical mechanics, condensed matter theory and string theories. They also play an important role in our present day understanding of quantum field theories.  Quantum field theories can be thought of as conformal field theories deformed by some relevant perturbations. In that sense CFTs are very special points in the space of all quantum field theories. CFTs appear in physical systems when there is no characteristic length scale. Therefore the correlation functions can only have power laws. It was pointed out in the early 90's \cite{Rozansky:1992rx, Saleur:1991hk, Gurarie:1993xq} that the structure of general CFTs allows the presence of multiplicative logarithms in
  correlation functions even at an RG fixed point. Such theories are called logarithmic conformal field theories (LogCFTs).
  Since CFTs come with no length/mass scale, one may wonder how there can be logarithms in their correlation functions. The answer lies on non-diagonalizable action of some operators of the type of a Hamiltonian.
  %Nonunitary evolution" $e^{t H}$  applications to models with disorder, systems with transient and recurrent states (sand-pile model), percolation, etc. 
  Let's see via an illustrative toy example how logarithms can appear in a CFT. Suppose the dilatation operator $\mathcal{L}_0 \sim x \frac{\partial}{\partial x}$ acts non-diagonally on a two component scalar $ \mathbb{G}$,
  \begin{align}
  \mathcal{L}_0 \mathbb{G} = \left[ 
  \begin{array}{cc}
  \Delta & 0 \\
  1 & \Delta \\
  \end{array}
  \right]
  \left[ 
  \begin{array}{c}
  g_1(x) \\
  g_2(x)\\ \end{array}
  \right]
  \end{align}
  where $\mathcal{L}_0 = [x \frac{\partial}{\partial x}, x \frac{\partial}{\partial x}]$ and $\mathbb{G} = \left[
  \begin{array}{c}
  g_1(x) \\
  g_2(x)\\ \end{array}
  \right]$. From this non-diagonal action of $\mathcal{L}_0$ we get the following equations,
  \begin{align} 
  x \,g_1'(x) &= \Delta \, g_1(x) \\ 
  x\, g_2'(x) &= \Delta \, g_2(x) + g_1(x)\,.
  \end{align}
  These are two first order differential equations with the following solutions,
  \begin{align}
  g_1 (x) &= B \,x^\Delta \\
  g_2 (x) &= A \,x^\Delta + B  \, x^\Delta \log(x)\,.
  \end{align}
  From representation theory perspective these are \emph{irreducible} but  \emph{indecomposable} representation of conformal group. This implies that the transformation of a two component scalar of dimension $\Delta$ involves a logarithm in the correlation function. In general there can be logarithmic multiplets of rank $r \geq 1$ which can be built by acting on $r$ primary states $|\mathcal{O}_i \rangle$ for $i=1, \cdots r$ obeying the highest-weight condition,
 \be
 K_{\mu} | \mathcal{O}_i \rangle =0 \, ,
 \ee
 where $K_{\mu}$ is the generator for the special conformal transformation. Hence a LogCFT consists of $r$-primaries and all their descendants.
 However, LogCFTs  are less constrained than unitary rational CFTs. %Therefore exploring properties of LogCFTs directly by  generalising  notions from ordinary CFTs are not very useful. 
 In \cite{Cardy:2013rqg}  Cardy took a complementary approach considering LogCFTs as limits of ordinary, non-logarithmic  CFTs, whose physical
  interpretation is already well understood, as a parameter is taken to a particular value.
  In this way the properties of the corresponding LogCFTs can be derived  and we are able  to understand exactly where the logarithms should appear in the physical observables. While that approach was quite general, it was illustrated  with several examples of physical interest, including quenched random magnets, self-avoiding walks, and percolation which makes it evident that LogCFTs are not merely mathematical artifact. One of the earliest physical systems LogCFTs  made its appearance was  in the context of polymers and percolation \cite{Saleur:1991hk, Cardy:1999zp}.  Several works has been  done in other statistical mechanical systems \emph{e.g.} 2D turbulence \cite{RahimiTabar:1995dh, RahimiTabar:1996si, Flohr:1996ik}, and the abelian sand-pile model \cite{Ruelle:2002jy, Mahieu:2001iv, Jeng:2006tg}. There are also applications in quantum condensed matter systems like disordered \cite{Caux:1995nm, Gurarie:1999bp} and the quantum Hall systems \cite{Bhaseen:1999nm, Kogan:1999hz, Ludwig:2000em, Read:2001pz}. LogCFTs have also been studied in the context of worldsheet string theory \cite{Kogan:1995df, Periwal:1996pw, Mavromatos:1998nz, Sfetsos:2002cn, Bakas:2002qh, Kogan:2000nw} and  AdS/CFT correspondence \cite{Ghezelbash:1998rj, Kogan:1999bn, MoghimiAraghi:2001fg, JabbariFaruji:2002xz, Grumiller:2013at}.  Recently Witten \cite{Witten:2018xfj}  found the necessity  for  LogCFTs in a first principle string theory derivation of entanglement entropy.

Although the  existence and appearance of LogCFTs in many physical systems are known for a long time, very little has been explored particularly in higher dimensions (see \emph{e.g.} \cite{Hogervorst:2016itc}). In higher dimensional CFTs conformal bootstrap has been used very successfully to study the spectrum, chaos, etc.  The idea of conformal bootstrap is to constrain a theory by exploiting the underlying conformal symmetry. The four point function of a CFT can be studied by imposing crossing symmetry and this leads to non-trivial constraint on the spectrum of the theory. The study of conformal bootstrap was initiated in \cite{Migdal:1972tk, Ferrara:1973vz, Ferrara:1973yt, Polyakov:1974gs} which is revived in its modern incarnation in \cite{Rattazzi:2008pe} and subsequently studied in \cite{Rychkov:2009ij, Poland:2010wg, Rattazzi:2010gj, Rattazzi:2010yc, Poland:2011ey, Gliozzi:2013ysa, Gliozzi:2014jsa, ElShowk:2012ht, El-Showk:2014dwa, Kos:2016ysd}. For numerical bootstrap  unitarity (\emph{i.e.} positivity of OPE coefficients)  plays an important role. Thus within known numerical techniques it's not easy to implement conformal bootstrap to LogCFTs. On the other hand there are analytical methods namely large spin  bootstrap \cite{Fitzpatrick:2012yx, Komargodski:2012ek, Fitzpatrick:2014vua,  Alday:2015eya, Kaviraj:2015cxa, Kaviraj:2015xsa, Alday:2015ota, Alday:2015ewa, Li:2015rfa, Dey:2016zbg, Alday:2016njk, Alday:2016jfr, Dey:2017fab, Dey:2017oim} and Polyakov-Mellin bootstrap \cite{Gopakumar:2016wkt, Gopakumar:2016cpb, Dey:2016mcs, Gopakumar:2018xqi, Ghosh:2018bgd} where one doesn't necessarily require unitarity. This note is our first step toward exploring the plethora of interesting LogCFTs which appear in different branches of physics as mentioned above, using analytic bootstrap. We will also see that the constraints imposed by crossing symmetry on the four point correlator dictate the spectrum of the large spin sector of LogCFT. This is an universal sector and \emph{any} particular LogCFT should contain this at leading order in large spin. In this sense our computation can be considered as a natural starting point to explore any  LogCFT. In this article we use the conformal bootstrap techniques in the large spin limit to compute the anomalous dimension of logarithmic operators for a particular type of LogCFT. We also compute it (strictly speaking, for a  special case) from effective dual gravitational description where anomalous dimension of the double-trace operators is identified with the binding energy of two rapidly rotating particles inside AdS \cite{Fitzpatrick:2014vua}. By doing so we show that cluster decomposition principle holds true even for non-unitary LogCFTs as long as the scaling dimensions are positive. 
   %\textcolor{blue}{\bf Some more details about bootstrap can be added?} \\  
  
 The paper is organised as follows. In  section \ref{sec:LGFF} we introduce the concept of Logarithmic Generalised Free Field (LGFF) theory  and  perform the meanfield analysis for rank-2   logarithmic scalar correlators. Section \ref{anmg} describes how to compute the leading correction to the anomalous dimension of rank-3 even spin double trace logarithmic operators appearing in the spectrum. In section \ref{sec:Bulk} we independently compute the  anomalous dimension from a dual bulk effective theory.  We conclude with some forward-looking remarks about the possible future directions in section \ref{sec:Conc}. Appendices contain some details of the computation.

  \section{\label{sec:LGFF}Logarithmic generalized free fields}

 In this section we consider  a particular  four dimensional LogCFT (following  \cite{Hogervorst:2016itc} and the references therein) : the logarithmic counterpart of the well-known \emph{generalized free fields} (GFF). %, which is defined in any dimension $d$
 In particular, this is a theory of a rank-two scalar multiplet $\phi_i$ ($i=1, 2$) of dimension $\DD_\phi$ with the following two-point function,
 \beq
 \label{eq:GFF2pt}
 \expec{\phi_i(x) \phi_j(0)} = \frac{1}{|x|^{2\DD_\phi}} \begin{pmatrix} - \ln x^2 & 1 \\  1 & 0 \end{pmatrix}_{ij}\,.
 \eeq
  
  The operator content of the $\phi_i \times \phi_j$ OPE consists of the unit operator $\unit$ as well as a tower of ``double-trace'' primaries $ \Oo_{i j}^{(n, \ell)}$. For even $\l$ and fixed $n$ there are three different double-trace operators, whereas for odd $\l$ there is only one which form a rank-3 and rank-1 multiplet respectively. 
 % The even-spin primaries  form rank-three multiplets in the logarithmic GFF. 
 For the simplest case with $n=0$ scalar  this logarithmic triplet is defined by the following operators, 
  \beq
  \label{eq:scalar triplet}
  S_1 = \frac{1}{2} \NO{(\phi_1)^2}\,, \quad S_2 = \NO{\phi_1 \phi_2}\,, \qquad S_3 = \NO{(\phi_2)^2}
  \eeq
  where $\NO{\phantom{ }}$ denotes normal ordering. Now we would like to bootstrap  the four point correlator  of four identical scalar rank-2 logarithmic operators .

 Let us begin with a brief review of  how the bootstrap equation can be used to reproduce the results from the mean field theory correlator in the usual (or non logarithmic) CFT. Consider a four-point correlator of four identical scalar operators $\phi$ of dimension $\Delta_\phi$. 
 This correlator has a conformal block decomposition is $s$ and $t$ channel.
  \begin{align}\label{beq}
  \langle \phi(x_1)\, \phi(x_2)\, \phi(x_3)\, \phi(x_4)\rangle  &= \frac{1}{\left(x_{12}\,x_{34}\right)^{2\Delta_\phi}}\bigg(1+\bigg(\frac{u}{v}\bigg)^{\Delta_\phi}+u^{\Delta_\phi}\bigg)\nn
  &=  \frac{1}{\left(x_{12}\,x_{34}\right)^{2\Delta_\phi}} \sum_{\Delta, \ell} c_{\Delta, \ell}\, \mathcal{G}_{\Delta, \ell}(u, v)\nn
  &= \frac{1}{\left(x_{13}\,x_{24}\right)^{2\Delta_\phi}} \sum_{\Delta, \ell} c_{\Delta, \ell}\, \mathcal{G}_{\Delta, \ell}(v, u),
  \end{align}
 where $x_{i j}= x_i-x_j$ and the conformal cross ratios are defined as,
 \begin{align}
 & u= \frac{x_{12}^2\,x_{34}^2}{x_{13}^2\,x_{24}^2}= z\, \zb , \qquad v= \frac{x_{14}^2\,x_{23}^2}{x_{13}^2\,x_{24}^2}= (1-z)(1-\zb)\,.
  \end{align}
  We will interchangeably use $u, v$ or $z, \zb$.  
Here $\Delta, \ell, c_{\Delta, \ell}$,  are the dimension, spin and OPE coefficients of the operators getting exchanged in the intermediate channel and $\mathcal{G}_{\Delta, \ell}(u, v)$ are the conformal blocks,
  % \begin{align}\label{block2}
  %&q_{\ell, n} = \frac{2^{\ell} \left(\left(-h+\Delta _1+1\right)_n \left(-h+\Delta _2+1\right)_n \left(\Delta _1\right)_{l+n} \left(\Delta _2\right)_{l+n}\right)}{l! n! (h+l)_n \left(-2 h+n+\Delta _1+\Delta _2+1\right)_n \left(l+2 n+\Delta _1+\Delta _2-1\right)_l \left(-h+l+n+\Delta _1+\Delta _2\right)_n}\nn
  %&G_{\DD, \ell}(u, v, \DD_i) = u^{-\frac{1}{6} ({\Delta_1}+{\Delta_2}+{\Delta_3}+{\Delta_4})}\, v^{\frac{1}{6} (-{\Delta_1}+2 {\Delta_2}+2 {\Delta_3}-{\Delta_4})}\, \hat{G}_{\DD, \ell}(u, v)\,,
  %\end{align}
  \be\label{block}
  \mathcal{G}_{\DD, \ell}(u, v) = \frac{1}{2^{\ell}}\frac{z\, \bz}{z- \bz}\bigg(k_{2\ell+\tau}(z)\,k_{\tau-2}(\bz)-k_{2\ell+\tau}(\bz)\,k_{\tau-2}(z)\bigg)\,,
  \ee
  with,
  \be
  k_{\beta}(x)=x^{\beta /2} \, _2F_1\left(\frac{\beta }{2},\frac{\beta }{2};\beta ;x\right)\,.
  \ee
  % The cross ratios are defined as,
%  \begin{align}
 %& u= \frac{x_{12}^2\,x_{34}^2}{x_{13}^2\,x_{24}^2}= z\, \zb , \qquad v= \frac{x_{14}^2\,x_{23}^2}{x_{13}^2\,x_{24}^2}= (1-z)(1-\zb)\,.
  %\end{align}
  %We will interchangeably use $u, v$ or $z, \zb$.  
  The equality of the first and the third line of \eqref{beq} reads,
  %\begin{align}
  %&\sum_{\Delta, \ell} c_{\Delta, \ell}\, \mathcal{G}_{\Delta, \ell}(u, v) =\bigg(\frac{u}{v}\bigg)^{\Delta_\phi} \sum_{\Delta, \ell} c_{\Delta, \ell}\, \mathcal{G}_{\Delta, \ell}(v, u)
 % \end{align}
 % Note that the conformal blocks $\mathcal{G}_{\Delta, \ell}(u, v) $ can be expanded around $u=0$ as follows,
  \be\label{mft2}
  1+ v^{-\Delta_\phi} + u^{-\Delta_\phi}= v^{-\Delta_\phi}\sum_{\Delta, \ell} c_{\Delta, \ell}\, \mathcal{G}_{\Delta, \ell}( v, u).
  \ee
  In mean field theory the intermediate operators consist of the identity and the double-trace operators $\mathcal{O}_{n, \ell}$ having dimension $\Delta= 2\DD_{\phi}+2n+\ell$. Now let us focus on \eqref{mft2} in the limit $u \sim 0$.  The conformal blocks $\mathcal{G}_{\Delta, \ell}(v, u) $ can be schematically expanded around $u \sim 0$ as follows,
  \begin{align}
 v^{-\Delta_\phi}  \mathcal{G}_{2\Delta_\phi+2n, \ell}(v, u) \sim \sum_{m=0}^{\infty}u^m\, f_m(v)+ \log u \sum_{m=0}^{\infty}u^m\, \tilde{f}_m(v).
  \end{align}
  Note that the lhs of \eqref{mft2} has a power law singularity as $u^{-\Delta_\phi}$ whereas each term on the rhs has a
  $\log u$ singularity.  Hence the power law singularity can not be reproduced by summing over a finite number of terms on the rhs. It can be shown
  \cite{Fitzpatrick:2012yx, Komargodski:2012ek} that by summing over large spin $\ell \gg 1$ operators in the rhs we can reproduce the lhs of \eqref{mft2}. This indicates why the large spin operators are important to reproduce the mean field theory of an ordinary CFT.
  
    Now we will discuss the importance of large spin operators in the context of LogCFT following the same steps as above. We consider the four-point function of a rank two scalar operator $\phi_i$  of dimension $\DD_\f${\footnote{For simplicity we consider only rank two operators. However, this can be generalised to higher rank operators as well.}}. 
    %%%%%%%%%%%%%%%%%%%%%%%%%%%%%%%%%%%%%%%%%%%%%%%%
    It was shown in \cite{Hogervorst:2016itc} that the four point function can be written as,
    \begin{align}
    \langle \phi_i(x_1)\, \phi_j(x_2)\, \phi_k(x_3)\, \phi_{\ell}(x_4)\rangle =  F_{i j k \ell} (u, v, \zeta_m)\, P_{\Delta_\phi \Delta_\phi \Delta_\phi \Delta_\phi}(x_n)
    \end{align}
    where, 
   \be
     P_{\Delta_1 \Delta_2 \Delta_3 \Delta_4}(x_n) = \prod _{n< r} \frac{1}{|x_{n r}|^{\DD_n-\DD_r-\Sigma/3}}, \qquad \Sigma := \sum_{i=1}^{4} \DD_i\,,
    \ee 
    and,
    \be
    \zeta_m= \partial_{\DD_m} \ln P_{\Delta_1 \Delta_2 \Delta_3 \Delta_4}(x_n)\,.
    \ee
    $F_{i j k \ell} $ is a tensor which must satisfy the cyclic permutation symmetry under the exchange of $(x_1, x_2, x_3, x_4) \rightarrow (x_2, x_3, x_4, x_1)$ combined by the exchange $x_1 \leftrightarrow x_2$,
    \be
    F_{i j k \ell} (u, v, \zeta_1,  \zeta_2,  \zeta_3,  \zeta_4)=  F_{ j k \ell i} (v, u, \zeta_2,  \zeta_3,  \zeta_4,  \zeta_1)=  F_{ j i k \ell} (u/v, 1/v, \zeta_2,  \zeta_1,  \zeta_3,  \zeta_4)\,.
    \ee
   Following the steps mentioned in \cite{Hogervorst:2016itc} it can be shown that the constraints imposed by conformal invariance  on $F_{i j k \ell} $  allows this to be written in terms of five conformally invariant functions $\mathcal{F}_{\alpha} (u, v), \alpha=1,\cdots 5$ as follows,
    \begin{align}
    F_{1111} & = \mathcal{F}_{1} (u, v)+ \sum_{i}\zeta_i \, \mathcal{F}_{2} (u, v) +(\zeta_1\, \zeta_2+\zeta_3\, \zeta_4)\, \mathcal{F}_{3} (u, v) \nn
     & + (\zeta_1\, \zeta_3+\zeta_2\, \zeta_4)\, \mathcal{F}_{3} (1/u, v/u)+ (\zeta_1\, \zeta_4+\zeta_2\, \zeta_3)\, \mathcal{F}_{3} (v,u) \nn
     &+ \sum_{i< j<k}\zeta_i\, \zeta_j\, \zeta_k\, \mathcal{F}_{4} (u, v) + \zeta_1\,\zeta_2\,\zeta_3\,\zeta_4\,\mathcal{F}_{5} (u, v),\nn
  F_{1112} & =\mathcal{F}_{2} (u, v)+ \zeta_1\, \mathcal{F}_{3} (v, u)+ \zeta_2\, \mathcal{F}_{3} (1/u, v/u) +\zeta_3\,  \mathcal{F}_{3} (u, v)\nn
  &+(\zeta_1\, \zeta_2+\zeta_1\, \zeta_3+ \zeta_2\, \zeta_3)\, \mathcal{F}_{4} (u, v)+ \zeta_1\,\zeta_2\,\zeta_3\, \mathcal{F}_{5} (u, v)\nn
 F_{1122} & = \mathcal{F}_{3} (u, v) +(\zeta_1+\zeta_2)\,\mathcal{F}_{4} (u, v) + \zeta_1\,\zeta_2\,\mathcal{F}_{5} (u, v) \nn
 F_{1222} & = \mathcal{F}_{4} (u, v) +\zeta_1\, \mathcal{F}_{5} (u, v) \nn
 F_{2222} & = \mathcal{F}_{5} (u, v)\,.
    \end{align}
   Hence the conformal bootstrap constraint on the correlator essentially reduces to the  following crossing symmetry condition on the functions $\mathcal{F}_{\alpha} (u, v)$, 
    \begin{align}
    & \mathcal{F}_{\alpha} (u, v) =  \mathcal{F}_{\alpha} (v, u) =  \mathcal{F}_{\alpha} (u/v, 1/v) \qquad{\rm{for}}  \qquad \alpha=1, 2, 4, 5 ;\nn
    & \mathcal{F}_{3} (u, v) = \mathcal{F}_{3} (u/v, 1/v) \,.
    \end{align}
  %  $\langle \phi_i(x_1)\, \phi_j(x_2)\, \phi_k(x_3)\, \phi_{\ell}(x_4)\rangle$ is given by five different functions $\mathcal{F}_{\alpha} (u, v), \alpha=1,\cdots 5$ . 
  Each of these functions can be decomposed into conformal blocks which results in five bootstrap equations for $\mathcal{F}_{\alpha}(u, v)$ .
 Let us  focus on the bootstrap equation for $\mathcal{F}_2(u, v)$,
  \be\label{boot}
  \mathcal{F}_2(u, v)=   \mathcal{F}_2(v, u)\,.
  \ee
   In this case  the exchange operators in each channel are  rank-$r$ operator  ($r=3$ for even spin and $r=1$ for odd spin) $\mathcal{O}_p$ of spin $\ell$   having the three-point function $\langle \phi_i \phi_j \mathcal{O}_p\rangle$ characterized by the following OPE coefficients $\lambda_{i j p}$, %for $p=1,\cdots r$,
 \begin{align}
& \lambda_{1 1 p}=a_p,\quad \lambda_{1 2 p}= \lambda_{2 1 p}= b_p \quad \lambda_{2 2 p}=c_p \quad {\rm{for \,\,even\, \,} \ell}, \quad p=1, 2 , 3 ,\nonumber\\
&\lambda_{1 1 p}= 0,\quad \lambda_{1 2 p}= -\lambda_{2 1 p}={\tilde{ b}}_p , \quad \lambda_{2 2 p}=0, \quad {\rm{for \,\,odd\, \,} \ell},\quad p=1 \,.
 \end{align}
  The conformal block decomposition %of four external saclars  $\phi_i$ of dimension $\Delta_i$
   for $\mathcal{F}_2(u, v)$ is given by summing over even and odd spin rank-$r$ operators $\mathcal{O}_p$ with dimension $\Delta$ and spin $\ell$,
  \be
 \mathcal{F}_2(u, v) =\sum_{\mathcal{O}} \mathcal{D}_{\mathcal{O}}\, G_{\DD, \ell}(u, v, \DD_i)
  \ee
  where, 
  \begin{align}\label{vpq}
  & \mathcal{D}_{\mathcal{O}}= \sum_{p,q = 1}^{r} [a_p+ b_p (\partial_{\Delta_1}+\partial_{\Delta_2})+ c_p \,\partial_{\Delta_1} \,\partial_{\Delta_2}] [b_q+ c_q\,\partial_{\Delta_3}]V^{pq}(\partial_{\mathcal{O}})\quad {\rm{for \,\,even\, \,} \ell}\,,\nonumber\\
  & \mathcal{D}_{\mathcal{O}}= \sum_{p,q = 1}^{r}{\tilde{b}}_p\,{\tilde{b}}_q(\partial_{\Delta_2}-\partial_{\Delta_1})V^{pq}(\partial_{\mathcal{O}})\quad {\rm{for \,\,odd\, \,} \ell}\,,\nonumber\\
  & V^{pq}(\partial) = 
    \begin{cases}
    \partial^n/n!, \quad \text{if } \,\,n = p+q-r-1\geq 0 \\
      0, \quad \quad \quad  \text{if }\, n < 0
     \end{cases}
  \end{align}
  and the $ G_{\Delta, \ell}$'s are defined in \eqref{block2}.
  For identical scalars the odd spin contribution vanishes because of the following identity (see \cite{Dolan:2000ut,Dolan:2003hv,Dolan:2011dv}),
  \begin{align}
  &\frac{\partial}{\partial_{\Delta_1}}G_{\DD, \ell}(u, v, \DD_i)\bigg|_{\DD_i=\DD_{\phi}}=
  \cdots = \frac{\partial}{\partial_{\Delta_4}}G_{\DD, \ell}(u, v, \DD_i)\bigg|_{\DD_i=\DD_{\phi}}= \frac{1}{12}\log\bigg( \frac{v}{u^2}\bigg)G_{\DD, \ell}(u, v, \DD_i)\bigg|_{\DD_i=\DD_{\phi}}.
  \end{align}
  %For identical scalars $\Delta_i= \Delta_{\phi}$.
  The four point function in the mean field theory can be computed using Wick's theorem and is given by  \cite{Hogervorst:2016itc},
  \be\label{mft}
    \mathcal{F}_2(v, u)= \frac{1}{6} \bigg[\left(\frac{v}{u^2}\right)^{\Delta_\phi /3} \log \left(\frac{v}{u^2}\right)+\left(\frac{u}{v^2}\right)^{\Delta_\phi /3} \log \left(\frac{u}{v^2}\right)+(u\, v)^{\Delta_\phi /3} \log (u \,v)\bigg]\,.
  \ee
 In this case the operators $\mathcal{O}_{n, \ell}$ have known OPE coefficient $q_{_{n, \ell}}$ \cite{Fitzpatrick:2011dm},
\be
q_{_{n, \ell}} = \frac{2^{\ell} \,\left(\Delta _1+1-h\right)_n \left(\Delta _2+1-h\right)_n \left(\Delta _1\right)_{\ell+n} \left(\Delta _2\right)_{\ell+n}}{\ell! n! (h+\ell)_n \left(\Delta _1+\Delta _2+n+1-2h\right)_n \left(\ell+2 n+\Delta _1+\Delta _2-1\right)_{\ell} \left(\Delta _1+\Delta _2+\ell+n-h\right)_n}
\ee
where $h=d/2= 2 $, $(a)_b=\frac{\Gamma(a+b)}{\Gamma(b)}$ . %and $d$ is the space-time dimension.
The $s$-channel decomposition of the correlator is given by,
 \be\label{sch}
    \mathcal{F}_2(u, v)=\frac{1}{6}\bigg({\frac{v}{u^2}}\bigg)^{{\DD_\f}/{3}} \log\bigg(\frac{v}{u^2}\bigg)+\sum_{\ell, n=0}^{\infty}{ {\mathcal{D}}^{(n, \ell) }}\,G_{\DD, \ell}(u, v, \DD_i)  \bigg|_{\DD_i=\DD_\f, \DD=2\DD_\f+2n+\ell}
 \ee
 with,
 \be\label{der}
 { {\mathcal{D}}^{(n, \ell) }}= \partial_{\DD_\f} q_{_{n, \ell}}+ q_{_{n, \ell}} (2 \partial_{\DD} +\partial_{\DD_1}+\partial_{\DD_2}+2 \partial_{\DD_3})\,,
 \ee
 \begin{align}\label{block2}
%&q_{\ell, n} = \frac{2^{\ell} \left(\left(-h+\Delta _1+1\right)_n \left(-h+\Delta _2+1\right)_n \left(\Delta _1\right)_{l+n} \left(\Delta _2\right)_{l+n}\right)}{l! n! (h+l)_n \left(-2 h+n+\Delta _1+\Delta _2+1\right)_n \left(l+2 n+\Delta _1+\Delta _2-1\right)_l \left(-h+l+n+\Delta _1+\Delta _2\right)_n}\nn
&G_{\DD, \ell}(u, v, \DD_i) = u^{-\frac{1}{6} ({\Delta_1}+{\Delta_2}+{\Delta_3}+{\Delta_4})}\, v^{\frac{1}{6} (-{\Delta_1}+2 {\Delta_2}+2 {\Delta_3}-{\Delta_4})}\, \mathcal{G}_{\DD, \ell}(u, v)\,.
\end{align}
For simplicity we will  focus on the double trace operators  $\mathcal{O}_{0, \ell}$ with $n=0$. The bootstrap equation \eqref{boot} is in general quite complicated to solve and we will see that it simplifies in the limit $v \ll  u  \ll 1$ and $\ell \gg 1$ . In this limit we approximate
$
\bz \sim 1-v, z \sim u
$\, and the leading term in \eqref{mft} is given by,
\be\label{mftrhs}
\mathcal{F}_2(v, u) \sim  \frac{1}{6}   \left(\frac{u}{v^2}\right)^{\Delta_\phi /3}\,\log \bigg(\frac{u}{v^2}\bigg).
\ee 
In \eqref{sch} the first term is the contribution from the identity operator exchange.
We will now use \eqref{boot} to reproduce \eqref{mftrhs} from the $s$- channel of \eqref{boot}\,. In the $\ell \gg 1$ limit the OPE coefficient and its derivative can be approximated as follows,
\be
q_{_{0, \ell}} \sim \frac{\sqrt{\pi } \,2^{-2 \Delta_\phi - \ell+2} \,{\ell}^{2 \Delta_\phi -\frac{3}{2}}}{\Gamma^2(\Delta_\phi )}\,,
\ee
\be
\partial_{\DD_\f} q_{_{0, \ell}}  \sim -\frac{\sqrt{\pi } \,2^{-2 \Delta_\phi - \ell+3} \ell^{2 \Delta_\phi -\frac{3}{2}}}{\Gamma^2(\Delta_\phi )} \left(\psi (\Delta_\phi )-\log \ell+\log 2\right)\,,
\ee
where $\psi$ is the digamma function. For large $\ell$ and $n=0$ we can approximate \eqref{sch} by,
\begin{align}\label{mftlhs}
 \mathcal{F}_2(u, v)%&=\sum_{\ell \gg 1, n=0}^{\infty}{ {\mathcal{D}}^{(\ell, n) 1112}}\,G_{\DD, \ell}(u, v, \DD_i) |_{\DD_i=\DD_\f, \DD=2\DD_\f+\ell}\nn
 & \sim \beta \sum_{\ell \gg 1} -\frac{4^{2} \,\ell^{2 \Delta_\phi -1} u^{\Delta_\phi }\, (\psi (\Delta_\phi )+\log (2) -\log \ell) }{\Gamma^2(\Delta_\phi )} K_0\left(2 \ell \sqrt{v}\right)\nn
&\quad + \beta\, \log u\, \sum_{\ell \gg 1}\frac{2^{3-2 \Delta_\phi } \,{\ell}^{2 \Delta_\phi -1}\, u^{\Delta_\phi } }{\Gamma^2(\Delta_\phi )} K_0\left(2 \ell \sqrt{v}\right)\nn
 &\quad+ \beta\, \frac{1}{3}\, \log \bigg(\frac{v}{u^2}\bigg)\,\sum_{\ell \gg 1}\frac{2^{3-2 \Delta_\phi } \,\ell^{2 \Delta_\phi -1}\, u^{\Delta_\phi } }{\Gamma^2(\Delta_\phi )} K_0\left(2 \ell \sqrt{v}\right),
\end{align}
where,
\be
\beta = u^{-{2 \Delta_\phi }/{3}} \,v^{\Delta _\phi /3}\,,
\ee
and $K_0$ is a modified Bessel function of the second kind.
We can approximate the large $\ell$ sum by an integral $\displaystyle \sum_{\ell \gg 1} \rightarrow \half\,\int \,d\ell$ where the $\half$ indicates that we are summing over only even spin operators. The integrals we need are  of the following kind (see  \cite{Alday:2007mf, Alday:2013cwa} for similar analysis), 
\begin{align}\label{int3}
&\int_{\ell_0}^{\infty} \ell^{2\DD_\f-1-a}\, K_0(2\ell\,\sqrt{v})\,\log \ell =- \frac{v^{-\DD_\f+a/2}\,\Gamma^2(\DD_\f-a/2)}{8} \bigg(\log v-2\, \psi (\Delta_\phi-a/2 )\bigg)+\cdots,\nn
&\int_{\ell_0}^{\infty} \ell^{2\DD_\f-1-a}\, K_0(2\ell\,\sqrt{v})= \frac{1}{4}v^{-\frac{a}{2}-\Delta_\phi } \Gamma^2\left(\frac{a}{2}+\Delta_\phi \right) +\cdots
\end{align}
where the dots denote the subleading terms in $v$. Using this  in \eqref{mftlhs} we get,
\begin{align}
\mathcal{F}_2(u, v)%& \sim \half\, \bigg(\frac{ u}{v^2}\bigg)^{\Delta_\phi /3} \bigg[- \log (4 v)+ \frac{1}{3} \log  \bigg(\frac{v}{u^2}\bigg) +2 \log 2+\log u \bigg] \nn
&\sim  \frac{1}{6}\bigg(\frac{ u}{v^2}\bigg)^{\Delta_\phi /3} \log \bigg(\frac{u}{v^2}\bigg).
\end{align}

 Thus we see that the leading behavior of $t$-channel in the limit $v \ll u \ll 1$ is reproduced by summing over large spin double trace operators in the $s$-channel of \eqref{boot}\,.

 The above result may seem like a mere consistency check. But it is worth noting that, just like GFF, given a primary operator $\mathcal{O}$ with twist $\tau$ in a LogCFT at large spin there must exist an infinite tower of primaries with twist $ 2\tau + n$ where $n =0, 1, 2, \ldots$ in order to satisfy crossing symmetry. Since this is just consequence of crossing symmetry this large spin spectrum is universal \emph{i.e.} independent of particular LogCFT. In this sense, the starting point of large spin analysis for any given LogCFT should be identical to this section.  

 \section{Computing the  anomalous dimensions at large $\ell$}\label{anmg}
 In this section we will use the bootstrap equation \eqref{boot} to compute the anomalous dimension of the large spin operators $\mathcal{O}_{0, \ell}$  in an interacting LogCFT with the following dimension,
 \be
 \DD= 2\DD_\f+\ell+\gamma_{_{0, \ell}}\,.
 \ee
  In order to do so we focus on the subleading corrections to \eqref{boot} in the $v \ll u \ll 1$ limit.   To compute the anomalous dimension we need to focus on the coefficient of $\gamma_{_0, _\ell}\,u^{\DD_\f}\log u$ from \eqref{sch}. 
  \be\label{logs}
  \mathcal{F}_2(u, v)\bigg|_{\log u} \sim \sum_{\ell \gg 1} { {\mathcal{D}}^{(n, \ell)}}\,G(u, v,\DD, \DD_i)  \bigg|_{\log u}.
  \ee

  At this point we assume that $\gamma_{0, \ell}$ has the following expansion in the $\ell \gg 1$ limit,
  \be\label{exp}
\gamma_{_{0, \ell}} \sim \frac{\gamma_{_0}}{\ell^a} +\cdots\,,
  \ee
  where the dots denote the subleading terms in $\ell$\,.  We have to determine the constant $a$ and the $\ell$ independent piece $\gamma_0$ from the bootstrap equation.  To extract the $\log u$ term above, we have to use  the  integrals (see \eqref{int3}). This results in the following term from  \eqref{logs},
  \be\label{schlog}
  \mathcal{F}_2(u, v)\bigg |_{\log u} \sim  {\gamma_{_0}}\, u^{{\Delta _{\phi }}/{3}} v^{ \left(3 a-4 \Delta _{\phi }\right)/6} \,\frac{  \Gamma^2\left(\Delta _{\phi }-\frac{a}{2}\right)}{ \Gamma^2\left(\Delta _{\phi }\right)}\,\bigg(\psi \left(\Delta _{\phi }-\frac{a}{2}\right)- \psi \left(\Delta _{\phi }\right)\bigg)\,.
  \ee

  Now we focus on the $t$-channel. The $t$-channel has an expansion controlled by the twist of the exchange operator and the subleading correction comes from the minimal twist operators  $\mathcal{O}_m$ of spin $\ell_m$, dimension $\DD_m$ and twist $\tau_m$\,. We assume that the operator dimensions are always positive and $\tau_m > 0$. We will focus on the coefficient of $\log u$ which comes from the $t$ channel and match it with \eqref{schlog}\,. We will use the following expansion for the $t$-channel conformal block \cite{Fitzpatrick:2012yx},
 \begin{align}
\mathcal{G}_{\tau_m, \ell_m}(v, u) \sim  C_m \, v^{\frac{\tau_m}{2}}\,(1-u)^{\ell_m}\, _2F_1\bigg(\frac{\tau_m}{2}+\ell_m, \frac{\tau_m}{2}+\ell_m, \tau_m+ 2\ell_m, 1-u\bigg).
 \end{align}
 At small $u$ we have ,
 \be
 _2F_1\bigg(\beta, \beta, 2\,\beta; 1-u\bigg) = \frac{\Gamma(2\beta)}{\Gamma^2(\beta)}\sum_{n=0}^{\infty} \bigg(\frac{(\beta)_n}{n!}\bigg)^n\, u^n\, \bigg[2\bigg( \psi(n+1)- \psi(\frac{\tau_m}{2}+\ell_m+n)\bigg)-\log u\bigg]\,.
 \ee

 To obtain the $\log u$ term  we will also need the expression for derivative of $\mathcal{G}(v, u)$ w.r.t $\Delta$ which is given by,
   \begin{align}
 \partial_{\DD} \mathcal{G}_{\DD, \ell}(v, u) = \frac{1}{2^{\ell}}\frac{(1-z) (1 - \bz)}{\bz- z}\bigg(  &\partial_{\DD} k_{2\ell+\tau}(1 - \bz)\,k_{\tau-2}(1-z) + k_{2\ell+\tau}(1 - \bz)\,  \partial_{\DD} k_{\tau-2}(1-z) \nn 
&- \partial_{\DD} k_{2\ell+\tau}(1-z)\,k_{\tau-2}(1 - \bz) - k_{2\ell+\tau}(1-z)\,  \partial_{\DD} k_{\tau-2}(1 - \bz) \bigg)\,.
 \end{align}
 A typical term in the above expression gives,
    \begin{align}
\partial_{_\DD} k_{\DD+\ell}(x)  = \frac{1}{2} \log x \ k_{\DD+\ell}(x) + x^{\frac{\DD+\ell}{2}} \,  \partial_{_\DD} \ _2F_1 \bigg(\frac{\DD+\ell}{2}, \frac{\DD+\ell}{2},\DD+\ell, x\bigg).
 \end{align}
 In the small $u$ limit, the $\log u$ terms can appear from the derivative of  hypergeometric pieces as follows,
     \begin{align}
 \partial_{\DD} \mathcal{G}_{\DD, \ell}(v, u)  \sim  \frac{1}{2^\ell} \, v^{\frac{\tau}{2}} (1-u)^{\frac{\tau}{2} + \ell} \, \partial_{_\DD} \ _2F_1 \bigg(\frac{\DD+\ell}{2}, \frac{\DD+\ell}{2},\DD+\ell, 1-u \bigg).
 \end{align}
 Similarly for $r^{th}$ derivative $\log u$ dependent terms can arise from,%{\bf PD: Check}
      \begin{align}
 \partial_{\DD}^r  \mathcal{G}_{\DD, \ell}(v, u)  \sim  \frac{1}{2^\ell} \, v^{\frac{\tau}{2}} (1-u)^{\frac{\tau}{2} + \ell} \, \partial_{_\DD}^r \ _2F_1 \bigg(\frac{\DD+\ell}{2}, \frac{\DD+\ell}{2},\DD+\ell, 1-u \bigg).
 \end{align}
 To collect the contribution to the coefficients of $\log u$ we need to use the integral representation of hypergeometric function,  \cite{NIST:DLMF}),
       \begin{align}
 \ _2F_1 \bigg(\frac{\DD+\ell}{2}, \frac{\DD+\ell}{2},\DD+\ell, 1-u \bigg) = \frac{\Gamma(\Delta+\ell)}{2 \pi i \,  \Gamma^4(\frac{\DD+\ell}{2}) } \int_{-i \infty}^{i \infty} \Gamma^2(\frac{\DD+\ell}{2} +t) \, \Gamma^2(-t)  \,u^t \,dt .
  \end{align}
 Since $u \ll 1$  we close the contour to the right hand side and pick residue from the double pole from $\Gamma^2(-t)$. It will produce two kind of terms :  $u^0$ (non-log term) and $u^0 \log u$ ($\log u$ term). 
 %\textcolor{blue}{(Explanation)} \\
 The relevant terms that can give $\log u$ terms in the $t$-channel are given by,
 \begin{align}
&\mathcal{D}_{\mathcal{O}} \, G(v, u) \bigg{|}_{relevant} \nn
& = \sum_{p,q=1}^{r} a_p \,b_q V^{pq} G(v, u) + (a_p \, c_q + 2 \,b_p\, b_q)\, \partial_1 V^{pq}\, G(v, u)\nn
&= \sum_{p,q=1}^{r} a_p \,b_q \left(\frac{u}{v^2}\right)^{\Delta_{\phi }/3} \bigg[\frac{\partial^m}{m!}  \mathcal{G}_{\DD, \ell}(v, u) \bigg]_{u^0 \log u} \nn
&+  \sum_{p,q=1}^{r} (a_p\, c_q + 2 \,b_p\, b_q) \frac{1}{12} (\log u) \, \left(\frac{u}{v^2}\right)^{\Delta_{\phi }/3} \bigg[\frac{\partial^m}{m!}  \mathcal{G}_{\DD, \ell}(v, u) \bigg]_{u^0}\,,
 \end{align}
 where  $m= p+q-r-1$ from \eqref{vpq}.
  Notice that all other terms will have $(\log u)^2, \, (\log u) (\log v)$ etc. Here we are interested in \emph{only} $\log u$ terms. For minimal twist operator exchange in t-channel,
  \begin{align}
 \partial_{\DD}^m  \mathcal{G}_{\tau_m, \ell_m}(v, u)  \sim  \frac{1}{2^\ell_m} \, v^{\frac{\tau_m}{2}} \bigg[I^m_{\log u} (\tau_m, \ell_m) + I^m_{non- \log u} (\tau_m, \ell_m) \bigg]
 \end{align}
 where,
 \begin{align}
 I^m_{\log u} (\tau, \ell)& =\partial_{\DD}^m \bigg(\frac{\Gamma(\Delta+\ell)}{  \Gamma^4(\frac{\DD+\ell}{2}) }\Gamma^2(\frac{\DD+\ell}{2} +t)\bigg) \bigg|_{t=0, \DD=\tau+\ell}\, , \nn
 I^m_{non- \log u} (\tau, \ell)&= 2\,\gamma_E\,\partial_{\DD}^m \bigg(\frac{\Gamma(\Delta+\ell)}{  \Gamma^4(\frac{\DD+\ell}{2}) }\Gamma^2(\frac{\DD+\ell}{2} +t)\bigg) \bigg|_{t=0,\, \DD=\tau+\ell}\nn
 &+\partial_{\DD}^m\, \partial_t \bigg(\frac{\Gamma(\Delta+\ell)}{  \Gamma^4(\frac{\DD+\ell}{2}) }\Gamma^2(\frac{\DD+\ell}{2} +t)\bigg) \bigg|_{t=0,\, \DD=\tau+\ell}\,.
 \end{align}
 Finally, the coefficient of $\log u$ term in the t-channel is given by,
  \begin{align}\label{tchlog}
  \mathcal{F}_2 (v, u) \bigg{|}_{\log u} &= \mathcal{D}_{\mathcal{O}} \, G(v, u) \bigg{|}_{\log u} \nn
 &= \sum_{p,q=1}^{r} a_p b_q \left(\frac{u}{v^2}\right)^{\Delta_{\phi }/3} \bigg[ \frac{1}{m!\, 2^{\ell_m}} v^{\tau_m/2} I^m_{\log u} (\tau_m, \ell_m) \bigg] \nn
 &+  \sum_{p,q=1}^{r} (a_p c_q + 2 b_p b_q) \frac{1}{12}\, \left(\frac{u}{v^2}\right)^{\Delta_{\phi }/3} \bigg[ \frac{1}{m!\, 2^{\ell_m}} v^{\tau_m/2} I^m_{non-\log u} (\tau_m, \ell_m) \bigg]\,.
 \end{align}

 Comparing \eqref{schlog} and \eqref{tchlog} we can see that the bootstrap equation is satisfied if $a=  \tau_m$\,. Clearly, this expansion  \eqref{exp} is valid only when $\tau_m >0$ which may not be true for any nonunitary theory where there is a possibility that $\Delta_m < 0$. Hence this is true only for a subsector of nonunitary theories where the dimensions are always positive. This results in the following anomalous dimension,
   \begin{align} \label{eq:anm-dim}
  \gamma_{_0} &= \frac{2 \Gamma^2(\DD_{\phi})}{\Gamma^2(\Delta_{\phi }-\frac{\tau_m}{2}) (- \psi(\Delta_{\phi }) + \psi(\Delta_{\phi }-\frac{\tau_m}{2}))}\nn %\bigg[\sum_{p,q}^{r} a_p b_q  \bigg[ \frac{1}{m!\, 2^{\ell_m}}  I^m_{\log u} (\tau_m, \ell_m) \bigg] \nn
  & \quad \times \sum_{p,q=1}^{r}   \frac{1}{m!\, 2^{\ell_m}} \bigg[ a_p \,b_q  I^m_{\log u} (\tau_m, \ell_m) + \frac{1}{12}{(a_p\, c_q + 2 \,b_p\, b_q)}\, I^m_{non-\log u} (\tau_m, \ell_m)  \bigg]\,.
  \end{align}
 % where $m= p+q-r-1$. 
 As long as there is no operator with negative scaling dimension in the theory, the anomalous dimension $\gamma_{_{0, \ell}} \sim \frac{\gamma_{_0}}{\ell^{\tau_m}}$ with $\tau_m > 0$ and in the strict $\ell \to \infty$ limit  $\gamma_{_{0, \ell}}$ vanishes. This suggests that cluster decomposition holds even for LogCFTs with positive scaling dimensions, which are necessarily non-unitary. In the next section we will see this has nice bulk interpretation as well. It means that two corresponding particles which are rapidly rotating in AdS space are well separated from each other.

 \section{\label{sec:Bulk}The dual gravity picture}
 
 The expression for the anomalous dimension derived above in \eqref{eq:anm-dim} is the main result of this note. Since there exists a holographic model dual to LogCFTs, it would be nice to see if the anomalous dimension can be interpreted (at least the parametric behaviour)   via holography. There are some works in the literature that study dual of LogCFTs \cite{Ghezelbash:1998rj,Kogan:1999bn,Bergshoeff:2012sc}. Here we consider a particular case of the main result, namely we consider a simplified setup where the minimal twist, $\tau_m = 2$. Consequently we can have a simple\footnote{Of course this won't reproduce the full parametric behaviour of the anomalous dimension. We leave that detailed computation for future work.} dual gravity description in the bulk. In this setup we try to compute the anomalous dimension of the exchanged operator for LogCFT derived above from dual classical gravity. We follow the formalism in  \cite{Fitzpatrick:2014vua} to obtain the anomalous dimension. It is known from the literature \cite{Ghezelbash:1998rj,Kogan:1999bn,Bergshoeff:2012sc}, dual gravitational effective theory of a LogCFT is given by some higher derivative EOM. \emph{E.g.} for a rank\emph{-r} LogCFT the dual scalar field in AdS satisfies the following EOM,
    \begin{align}
  (\square - M^2)^r \, \Phi^{(r)}_{\text{LCFT}} = 0.
  \end{align}
  In this paper we focus on rank\emph{-2} multiplet and therefore the EOM
  \begin{align}
  (\square - M^2)^2 \, \Phi_{\text{LCFT}} &= 0 \nn 
  (\square - M^2) \underbrace{(\square - M^2) \, \Phi_{\text{LCFT}}}_{ \Phi_{\text{CFT}}}&= 0 .
    \end{align}
  
  By definition $\Phi_{\text{CFT}}$ is the solution of KG equation in AdS$_{d+1}$\footnote{The bulk analysis is valid for arbitrary $d$. However in order to compare this with the LogCFT we will explicitly choose $d=4$.}
  \begin{align}\label{eq:metricGlobal}
  ds^2 = \frac{1}{\cos^2{\rho}} \bigg( -dt^2 + d \rho^2 + \sin^2{\rho}\, d\Omega_{d-1}^2 \bigg),
  \end{align}
   where $\rho = \frac{\pi}{2}$ is the boundary and we have taken the AdS radius to be one. We will be using global metric because the connection between fields in AdS and operators with definite scaling dimension in the CFT is more transparent in global coordinates than in Poinca\'{r}e  patch. The solution to the EOM is known  \cite{Balasubramanian:1998sn, Fitzpatrick:2010zm} and is given by,
   \begin{equation}
   \Phi_{\text{CFT}}(x)=\sum_{n, \ell, J} \phi_{n \ell J}(x) a_{n \ell J}+\phi_{n \ell J}^{*}(x) a_{n \ell J}^{\dagger}.
   \end{equation}
  where, 
    \begin{align}
  \phi_{n \ell J} &= \frac{1}{N_{\Delta, n, \ell}} \, e^{i \omega_{n, \ell} t} \, Y_{\ell J}(\Omega) \sin^\ell{\rho} \cos^\Delta{\rho} \,  _2F_1 \bigg(-n, \Delta + n + \ell, \ell+ \frac{d}{2}, \sin^2{\rho}\bigg) \label{eq:homoSol} \\
    \omega_{n, \ell} &\equiv \Delta + 2n + \ell , \quad M^2 \equiv \Delta (\Delta - d) \\
    N_{\Delta, n, \ell} &\equiv (-1)^n \sqrt{\frac{n! \, \Gamma^2(\ell+\frac{d}{2}) \, \Gamma(\Delta + n - \frac{d-2}{2})}{\Gamma( n + \ell + \frac{d}{2}) \, \Gamma(\Delta + n + \ell)}}
   \end{align}
   and $a_{n \ell J}^{\dagger}, a_{n \ell J}$ are creation and annihilation operators respectively.
  
  \subsection*{Solution to the bulk EOM}
  
  To obtain $\Phi_{\text{LCFT}}$ (which we call $\Phi$ now onward for brevity) one needs to solve the following differential equation
  \begin{align}\label{eq:inhomDE}
  (\square - M^2) \, \Phi =   \Phi_{\text{CFT}}
  \end{align}
  
  This is nothing but KG equation with a known source term $\Phi_{n, \ell, J}$ which is the solution to the homogeneous KG equation in AdS$_{d+1}$. The standard approach to solve such an inhomogeneous partial differential equation is by using Green function method. For that we need to solve the KG equation with a delta function source 
   \begin{align}
   (\square - M^2) \, G(x- x')  =   \delta^{d+1}(x - x') ,
   \end{align}
  then integrate that solution over the known function $\Phi_{\text{CFT}}$ 
   \begin{align} \label{eq:SolnGF}
 \Phi(x) = \int \, \sqrt{g} \, d^{d+1}x' \, G(x- x') \, \Phi_{\text{CFT}}(x').
   \end{align} 

The bulk-to-bulk propagator (Green functions) can be written as an infinite sum over the normalizable modes using the general Green's function formula \cite{Balasubramanian:1998sn}
\begin{align}\label{eq:GF}
i \, G (x, x') = \int \frac{d \omega}{2 \pi} \sum_{n', \ell', \vec{m}'} e^{i \omega (t - t')} \, \frac{\phi^*_{n, \ell, \vec{m}}(\vec{x}) \, \phi_{n, \ell, \vec{m}}(\vec{x}')}{\omega^2_{n', \ell'} - \omega^2},
\end{align}
where the normalizable modes $\phi_{n, \ell, \vec{m}}(x)$ are known functions,
\begin{align} 
\phi_{n, \ell, \vec{m}}(x) = \, Y_{\ell m}(\Omega) \sin^\ell {\rho} \cos^\Delta{\rho} \,  P_n^{\ell + d/2 -1, \nu}(\cos{2 \rho}),
\end{align}
where $P_n^{m, \nu}(\cos{2 \rho})$ are Jacobi polynomials. From \eqref{eq:homoSol}, \eqref{eq:SolnGF} and \eqref{eq:GF} we can get the solution for \eqref{eq:inhomDE} as follows (see appendix \ref{sec:int} for details), 
   \begin{align}
   \Phi(x) &= \int \sqrt{-g} \, d^{d+1}x' \, G(x, x') \, \Phi_{\text{CFT}}(x') \nn 
   &= \frac{Y^*_{\ell J}(\Omega) e^{i \omega_{n, \ell} t}}{i N_{\Delta, n, \ell}} \frac{n!}{(\ell +\frac{d}{2})_n} (\sin{\rho})^\ell (\cos{\rho})^\Delta \sum_{m=0}^{\infty} \frac{1}{(\omega^2_{m, \ell} - \omega^2_{n, \ell})}P_{m}^{\ell + d/2 -1, \nu}(\cos{2 \rho})  \nn
   & \quad \times \int_0^{\pi/2} d{\rho}'   (\sin{\rho'})^{2 \alpha +1} (\cos{\rho'})^{2 \beta -1}  P_{m}^{\alpha, \nu}(\cos{2 \rho'}) P_{n}^{\alpha, \nu}(\cos{2 \rho'}) \nn \label{eq:ansatz}
   &\equiv  \frac{Y^*_{\ell J}(\Omega) e^{i \omega_{n, \ell} t}}{i N_{\Delta, n, \ell}} \frac{n!}{(\ell + \frac{d}{2})_n} (\sin{\rho})^\ell (\cos{\rho})^\Delta \, f(\rho),
%   &= \int \, \sqrt{-g}  \, dt' d{\rho}' d{\Omega}' \, \int \frac{d \omega}{2 \pi i} \sum_{n, \ell, \vec{m}}  \, \frac{e^{i \omega (t - t')}}{\omega^2_{n, \ell} - \omega^2} \, \phi^*_{n, \ell, \vec{m}}(\vec{x}) \, \phi_{n, \ell, \vec{m}}(\vec{x}')\nn
%   & \quad \quad \times \frac{1}{N_{\Delta, n, \ell}} \, e^{i \omega_{n, \ell} t'} \, Y_{\ell J}(\Omega') \sin^\ell{\rho'} \cos^\Delta{\rho'} \,  _2F_1 \bigg(-n, \Delta + n + \ell, \ell + \frac{d}{2}, \sin^2{\rho'}\bigg).
   \end{align} 
   where $\sqrt{-g}  = (\sin{\rho'})^{-d-1}\, (\cos{\rho'})^{d-1}$, and $\alpha = \ell + \frac{d}{2} -1; \quad \beta = \Delta - \frac{d}{2} = \nu $. 
%Performing the $t'$, $\omega$ and $\Omega'$ integrals (see appendix \ref{sec:int} for details), we get
%  \begin{align} 
%\Phi(x) &= \int \, \sqrt{-g} \, d{\rho}' d{\Omega}' \, \sum_{n', l', \vec{m'}}  \, \frac{\phi^*_{n', l', \vec{m}'}(\vec{x}) \, \phi_{n', l', \vec{m}'}(\vec{x}')}{i (E^2_{n', l'} - E^2_{n,l})} \, \nn
%& \quad \quad \frac{ e^{i E_{n, l} t}}{N_{\Delta, n, l}} \, \, Y_{lJ}(\Omega') \sin^l{\rho'} \cos^\Delta{\rho'} \,  _2F_1 \bigg(-n, \Delta + n + l, l+ \frac{d}{2}, \sin^2{\rho'}\bigg) 
%\end{align} 
%
%Next we use the orthogonality of spherical harmonics (see \emph{e.g.} \cite{Cohen})
%   \begin{align} 
% \int  d \Omega' Y^*_{l' m'}(\Omega') Y_{l J}(\Omega')  = \delta_{l, l'} \delta_{J m'}
% \end{align}
%to obtain
%  \begin{align} 
%\Phi(x) &= \frac{Y^*_{\ell J}(\Omega) e^{i \omega_{n, \ell} t}}{i N_{\Delta, n, \ell}} \frac{n!}{(\ell +\frac{d}{2})_n} (\sin{\rho})^\ell (\cos{\rho})^\Delta \sum_{m=0}^{\infty} \frac{1}{(\omega^2_{m, \ell} - \omega^2_{n, \ell})}P_{m}^{\ell + d/2 -1, \nu}(\cos{2 \rho})  \nn
% & \quad \times \int_0^{\pi/2} d{\rho}'   (\sin{\rho'})^{2 \alpha +1} (\cos{\rho'})^{2 \beta -1}  P_{m}^{\alpha, \nu}(\cos{2 \rho'}) P_{n}^{\alpha, \nu}(\cos{2 \rho'}) \nn \label{eq:ansatz}
% &\equiv  \frac{Y^*_{\ell J}(\Omega) e^{i \omega_{n, \ell} t}}{i N_{\Delta, n, \ell}} \frac{n!}{(\ell + \frac{d}{2})_n} (\sin{\rho})^\ell (\cos{\rho})^\Delta \, f(\rho),
%\end{align} 
%where $\sqrt{-g}  = (\sin{\rho'})^{-d-1}\, (\cos{\rho'})^{d-1}$, and $\alpha = \ell + \frac{d}{2} -1; \quad \beta = \Delta - \frac{d}{2} = \nu $. 

To obtain the solution $\Phi(x)$ one needs to compute $f(\rho)$ by performing the $m$ sum and $\rho'$ integral. The alternative way would be to use \eqref{eq:ansatz} as an ansatz to the differential equation \eqref{eq:inhomDE}, and solve for $f(\rho)$. Pursuing the latter route we end up with the  following differential equation 
\begin{align}
f''(\rho ) - \big( (2 \Delta -3) \tan \rho - (2 \ell  + 3) \cot \rho \big) f'(\rho ) - (\Delta -\omega +\ell ) (\Delta +\omega +\ell ) f(\rho )  =\sec ^2 \rho, 
\end{align}
which can be solved exactly (see appendix \ref{sec:int} for details) to get the solution
\begin{align}
 f(\rho )  &=  - \gamma_{_E}   \, \frac{(-1)^n\,\Gamma (\Delta -2) \Gamma (n+\ell +2)}{\Gamma (n+\Delta -1) \Gamma (\ell +2)} \, _2F_1\left(-n,n+\ell +\Delta ;\ell +2;\sin ^2 \rho  \right).
\end{align}
Finally the full solution to \eqref{eq:inhomDE} is given by,
 \begin{align} 
\Phi(x) &=   - \gamma_{_E} \,  \frac{Y^*_{\ell J}(\Omega) e^{i \omega_{n, \ell} t}}{i N_{\Delta, n, \ell}}  (\sin{\rho})^\ell (\cos{\rho})^\Delta \frac{n!}{(\ell+\frac{d}{2})_n}  \, \frac{\Gamma (\Delta -2) \, \Gamma (n+ \ell +2)}{\Gamma (n+\Delta -1) \,  \Gamma (\ell + 2)} \nn
 & \quad \quad \times \, _2F_1\left(-n,n+\ell +\Delta ;\ell +2;\sin ^2 \rho  \right).
\end{align} 

\subsection*{The anomalous dimension as binding energy}
Since we know the solution to the dual bulk scalar field in AdS, following \cite{Fitzpatrick:2010zm,Fitzpatrick:2012yx,Fitzpatrick:2014vua} we can try to extract the anomalous dimension as binding energy of two-particle state. As we have mentioned before, global AdS is very useful to have such a  bulk interpretation because time translations  in global AdS are generated by the dilatation operator $D$ of the dual CFT, and therefore anomalous dimensions in the CFT are equivalent to energy shifts of bulk states due to interactions. 
%The picture works as follows. A scalar primary operator $\phi$ with dimension $\Delta_{\phi},$  creates a state $|\phi\rangle=\phi|0\rangle$  acting on the vacuum of the CFT. Now if  $\phi$ is a single-trace operator (at large $N$), then by AdS/CFT dictionary we can interpret $|\phi\rangle$ as a single-particle state in AdS. Let's consider the OPE of $\phi$  with itself,
%\begin{align} 
%\phi(x) \phi(0)=\sum_{\tau, \ell} c_{\tau, \ell} \,  f_{\tau, \ell}(x, \partial) \, \mathcal{O}_{\tau, \ell}(0),
%\end{align} 
%we can interpret the operators $\mathcal{O}_{\tau, \ell}$ as 2-particle states. Their anomalous dimensions are coming from the bulk interactions. Since we are focusing on large $\ell$ regime, the angular momentum of the two particles are very large and thus they are well separated inside AdS. 
%
%\begin{equation}
%\Phi(x)=\sum_{n, \ell, J} \Phi_{n \ell J}(x) a_{n \ell J}+\Phi_{n \ell J}^{*}(x) a_{n \ell J}^{\dagger}.
%\end{equation}
%The creation operator $a_{n l J}^{\dagger}$ acting on the bulk vacuum creates a one particle state 
%\begin{equation}  
%| \Phi ; n, \ell, J \rangle  \equiv a_{n \ell J}^{\dagger} | 0 \rangle .
%\end{equation}
%These states are in one-to-one correspondent to states created by a single trace operator $\mathcal{O}(0)$ acting at the origin of the dual CFT and its descendants.  

Here we calculate the first order shift in energy due to (Newtonian) gravitational interaction between the two rapidly orbiting particles. To simplify the computation, following \cite{Fitzpatrick:2014vua}, we fix one of them at the center of AdS$_5$\footnote{Notice that the bulk can be of arbitrary spacetime dimensions. We are choosing AdS$_5$ because in the CFT side we have used $4d$ conformal block expressions. Presumably the whole setup will go through for any $d \ge 3$.} and the other one will be moving with an effective large \emph{orbital} angular momentum $\ell_{orb}$. This is equivalent to studying orbital motion of a massive particle in AdS$_5$-Schwarzschild black hole.     

%
%\begin{figure}
%	\centering
% \includegraphics[width=10cm]{bulk-pic}
%  \caption{The left figure shows the bulk picture of two orbiting particles in AdS$_5$. For simplicity we can effectively think of that as one `heavy' particle is at rest at the centre of AdS and the other one orbiting around it with angular momentum $\ell_{orb}$. This amounts to the particle orbiting around AdS$_5$-Schwarzschild that has been depicted in the figure at the right.  }
%   \label{fig:bulk-pic}
%\end{figure}

\begin{equation}
d s^{2}=N(r) d t^{2}-\frac{1}{N(r)} d r^{2}-r^{2}  d \Omega_3^{2}
\end{equation}
with, 
\begin{equation}
N(r)=1-\frac{\mu}{r^{2}}+\frac{r^{2}}{R_{\mathrm{AdS}}^{2}},
\end{equation}
 and the mass of the BH is given by, $M_{_{BH}}=\frac{3 \, \text{vol}{(S^3)}}{16 \pi G_{_N}} \mu$. 
Note that the coordinate $r$ of this metric is related to \eqref{eq:metricGlobal} via the map : $r = \tan{\rho}$. The wave function in $r$ coordinate reduces to,
 \begin{align} 
 \Phi(x) &=  \underbrace{\left(\frac{1}{\epsilon }-\gamma_{_E} \right)\,  \frac{1}{i N_{\Delta, n, \ell_{o r b}}}  \frac{n!}{(\ell_{orb} + \frac{d}{2})_n}  \, \frac{\Gamma (\Delta -2) \, \Gamma (-\ell_{o r b} -1)}{\Gamma (n+\Delta -1) \,  \Gamma (-n-\ell_{o r b} -1)}}_{\mathcal{N}} \nn & \hspace{0.2cm} \, \times \, \underbrace{\frac{r^{\ell_{orb}}}{(1+r^2)^{\frac{\ell_{o r b}}{2}}}  \frac{1}{(1+r^2)^{\frac{\Delta}{2}}} \, _2F_1\left(-n,n+\ell_{o r b} +\Delta ;\ell_{o r b} +2; \frac{r^2}{1+r^2}  \right)}_{\mathcal{R}(r)} \nn 
 & \hspace*{4cm}\, \times \,  Y^*_{\ell J}(\Omega) \, e^{i \omega_{n, \ell_{o r b}} t}.
 \end{align}

We are interested in computing the energy shift of the orbiting particle due to gravitational attraction. In first oder in perturbation the shift in energy is given by, 
\begin{equation}
\begin{aligned} \delta E_{orb} &=\left\langle n, \ell_{orb}|\delta H| n, \ell_{orb}\right\rangle \\ &=-\frac{\mu}{4} \int d r r^{3} d^{3} \Omega\left\langle n, \ell_{o r b}\left|\left(\frac{r^{-2}}{\left(1+r^{2}\right)^{2}}\left(\partial_{t} \phi\right)^{2}+r^{-2}\left(\partial_{r} \phi\right)^{2}\right)\right| n, \ell_{orb}\right\rangle, \end{aligned}
\end{equation}

\begin{equation}
\delta E_{o r b}\left(n, \ell_{o r b}\right) = - \frac{\mu}{2} \, |\mathcal{N}|^2 \, \int r d r \left( \, \frac{1}{\left(1+r^{2}\right)^{2}} \, \omega_{\Delta n \ell_{o r b}}^{2}\left|\mathcal{R}(r)\right|^{2}+\left(\partial_{r} \mathcal{R}(r)\right)^{2}\right),
\end{equation}
where $\mathcal{R}(r)$ is the radial part of the wavefunction. We want to compute this binding energy and compare that with the anomalous dimension $\gamma_{_0}$ given in \eqref{eq:anm-dim}. Therefore it will suffice if we focus on the regime : $n=0$ and $\ell_{orb} \gg 1$, where our CFT computation is valid. In this limit, 
\begin{align}
\mathcal{R}(r) &= \frac{r^{\ell_{orb}}}{(1+r^2)^{\frac{\ell_{orb}}{2}}}  \frac{1}{(1+r^2)^{\frac{\Delta}{2}}}, \\
\mathcal{N} &=  i \gamma_{_E} \,  \sqrt{\frac{\Gamma (\ell_{orb} +\Delta )}{\Gamma (\Delta -1) \Gamma (\ell_{orb} +2)}}   \, \frac{\Gamma (\Delta -2)}{\Gamma (\Delta -1)}.
\end{align}
Performing the $r$ integral we find,
\begin{align}
\delta E_{o r b}\left(0, \ell_{o r b}\right) &=   - \frac{\mu}{2} \, |\mathcal{N}|^2 \, \bigg(-\frac{\Gamma \left(\Delta +\frac{3}{2}\right) (\Delta + \ell_{orb})^2 \Gamma \left(\ell_{orb} +\frac{1}{2}\right)}{2 \Gamma (\ell_{orb}+\Delta +2)} \nn
& \hspace*{4cm} + \frac{\Gamma \left(\Delta +\frac{1}{2}\right) (\Delta + \ell_{orb} ) ((4 \Delta +3) \ell_{orb} -\Delta ) \Gamma \left(\ell_{orb} -\frac{1}{2}\right)}{8 \Gamma (\ell_{orb} +\Delta +2)}\bigg) \nn 
&\approx  -  \, \gamma^2_{_E} \, \frac{2 G_{_N} \, M_{_{BH}}}{3 \pi} \frac{\Delta  (\Delta -1) \Gamma^2 (\Delta -2) }{\Gamma^2 (\Delta -1)}  \left(\frac{1}{\ell_{orb}}-\frac{1}{\ell^2_{orb}}\right).
\end{align}

Note that the formula we have is for a particle's motion around  the black hole. But originally we had two particle orbiting rapidly in vacuum AdS$_5$ as a the dual to double-trace primary operator. Therefore we need to map back to that two particle picture (see \cite{Fitzpatrick:2014vua}) to obtain,
%Notice that the orbit state has the same energy shift/binding energy as a primary with equal geodesic distance between the two particles, $d_{orb} = d_1 + d_2$.
%\begin{align}
%d_{1}=\frac{1}{2} \log \left(\frac{\ell_{\text {p}}}{\Delta_{1}}\right), \qquad
%d_{2}=\frac{1}{2} \log \left(\frac{\ell_{\text {p}}}{\Delta_{2}}\right)\,.
%\end{align}
%The geodesic distance from the centre of AdS to the orbiting particle (see \cite{Fitzpatrick:2014vua}) is given by,
%\begin{equation}
%\langle d \rangle \approx \frac{R_{\mathrm{AdS}}}{2} \log \left(\frac{2 \ell_{orb}}{\Delta}\right).
%\end{equation}
%Thus we can relate the orbital angular momentum and the `primary angular momentum' as follows,
%\begin{equation}
%\ell_{o r b}=\frac{\ell_{\text {p}}^{2}}{2 \Delta_{1}}.
%\end{equation}
%Now let's replace $M_{_{BH}} \approx \Delta_{1}$, $\Delta \equiv \Delta_{2}$ and $\ell_{\text {p}} \equiv \ell$, to get,
\begin{align}\label{eq:energy-shift}
\delta E_{o r b}\left(0, \ell_{o r b}\right)
&\approx   -  \, \gamma^2_{_E} \, \frac{2 G_{_N} \, \Delta_{1} }{3 \pi} \frac{\Delta_2  (\Delta_2 -1) \Gamma^2 (\Delta_2 -2) }{\Gamma^2 (\Delta_2 -1)}  \left(\frac{2 \Delta_1}{\ell^2} \right).
\end{align}
%We want the energy shift in the limit $\ell \gg \Delta_{1}, \Delta_{2} \gg 1$; in that limit
%\begin{align}
%\delta E_{o r b}\left(0, \ell_{o r b}\right)
%&\approx   -  \, \gamma^2_{_E} \, \frac{4 G_{_N} \, (\Delta_{1} \Delta_2)^2}{3 \pi} \left(\frac{1}{\ell^2} \right) \nn
%&= -   \, \gamma^2_{_E} \, \frac{(\Delta_{1} \Delta_2)^2}{6 \, c} \left(\frac{1}{\ell^2} \right) 
%\end{align}
%In the last line we have used the expression for central charge  $c = \frac{\pi}{8 G_{_N}}$ in spacetime dimensions $d=4$. \\
It is evident that \eqref{eq:energy-shift} doesn't identically match to \eqref{eq:anm-dim} that we derived using analytic bootstrap techniques. This is not unexpected, given the simpleminded dual gravity model for LogCFT we have considered. The only thing we want to extract from this result is the large $\ell$ dependence. The binding energy which is equivalent to the anomalous dimension in the LogCFT side decreases like $\frac{1}{\ell^2}$ since $\ell$ is large. In the $\ell \to \infty$ limit,  $\delta E_{o r b} \to 0$, which means the particles are far from each other in AdS space and therefore effectively behave as ``free" particles. This is a statement of cluster decomposition of the LogCFT in dual AdS language.

  \section{\label{sec:Conc}Conclusion}
In this paper we have studied logarithmic conformal field theory using analytic bootstrap techniques in four dimensions. In particular, we studied the four point correlator of rank-2 identical logarithmic scalars. Using the bootstrap techniques  we have shown how the intermediate double-trace operators in the large spin limit can reproduce the mean field theory correlator.  We have computed the leading correction to the anomalous dimension of even spin rank-3 operators appearing in the OPE of two rank-2 logarithmic scalars in the large spin limit. LogCFTs are known to have holographic dual. The anomalous dimension of  the double trace operators due to stress tensor exchange in four dimensions can be interpreted as the binding energy of the two particles rotating rapidly in global AdS. Our results also indicate that cluster decomposition holds (at least for large spin sector) even for certain class of non-unitary CFTs namely LogCFTs in $d >2$ with no operator with negative scaling dimensions. This is an interesting feature since unitarity condition plays key role in proving cluster decomposition for ordinary CFTs. %\textbf{Matching anomalous dimension from holography using higher rank stress tensor OPE.} \\
 There are many interesting directions to pursue in future. 
\begin{itemize}

\item
The subleading corrections in $1/\ell$ to the anomalous dimension can be computed. This can be simplified in Mellin space following \cite{Dey:2017fab}. One can also take into account the higher twist double trace operators for $n \neq 0$. It would be nice to compute the anomalous dimension in the bulk side due to the exchange of higher rank stress-tensor. Moreover, the bulk and boundary analysis of rank-2 LogCFTs may be generalised to rank-$r$ LogCFTs by incorporating higher derivative action in the dual  rank-$r$ LogCFTs. It would be interesting to repeat the analysis in general dimensions.

\item
In this note we have not used the mean field values of OPE coefficients $a_p,  b_p, c_p $ for the minimal twist operators in \eqref{eq:anm-dim}. Given a LogCFT one can compute these coefficients. As LogCFTs are necessarily non-unitary they can, in general, be complex numbers. This can lead to some interesting physical phenomena both in the LogCFT side and in the dual bulk picture (\emph{e.g.} the energy and loci of the rotating particles). 

\item
Probably the most physically interesting systems to study are those LogCFTs which appear in condensed matter systems. For example, it would be nice to explore the $Q$-state Potts model using conformal bootstrap. As we have mentioned before the large spin sectors of all these particular systems are identical and we have already analysed that in section \ref{sec:LGFF} in this paper. Also, in principle, the same large spin systemics should work for computing the corresponding anomalous dimensions for all those interesting LogCFTs. 
 
 \item
Another interesting but somewhat different direction will be to explore other known or conjectured non-unitary CFTs (which are not necessarily LogCFTs) by the same technique. E.g. one can analytically bootstrap non-unitary $\mathcal{N}=4$ theory  \cite{Vafa:2014iua}  in 4d with the gauge group $U(N+k|k)$. This theory is said to be indistinguishable from its unitary cousin namely $\mathcal{N}=4$ SYM with gauge group $SU(N)$ in arbitrary orders in $1/N$. Since bootstrap methods are non-perturbative one can hope to capture their difference which is expected to be $\mathcal{O}(e^{- N})$.

We hope to return to some of these problems in future. 
\end{itemize}

  \acknowledgments 
  We thank Agnese Bissi, John Cardy and Aninda Sinha for useful discussions. Specifically, we thank Agnese Bissi and Aninda Sinha for comments on the draft.
  PB is grateful to the MPI Partner group grant MAXPLA/PHY/2018577. PD is supported by the Knut and Alice Wallenberg Foundation grant KAW 2016.0129.

  \appendix

 \section{Conformal blocks at large $\ell$ }
   The conformal block in four dimensions is given by,
   \begin{align}\label{blockghat}
   {\mathcal{G}}_{\Delta, \ell}(u, v) %&= \frac{1}{2^{\ell}}\frac{z\, \bz}{z- \bz}\bigg(k_{2\ell+\tau}(z)\,k_{\tau-2}(\bz)-k_{2\ell+\tau}(\bz)\,k_{\tau-2}(z)\bigg)\nn
   & = \frac{1}{2^{\ell}}\frac{z\, \bz}{z- \bz}\bigg(k_{\DD+\ell}(z)\,k_{\DD-\ell-2}(\bz)-k_{\DD+\ell}(\bz)\,k_{\DD-\ell-2}(z)\bigg).
   \end{align} 
   We will be working in the limit $\ell \rightarrow \infty$ keeping $v\, \ell^2= y$ fixed. %In the lightcone limit we approximate $u \sim z$ and $v \sim 1-\zb$. We will use these cross-ratios interchangably.
    Note that $k_{2\ell+\tau}(z)$ is proportional to $z^{\ell}$ and since we are in the regime $z \ll 1$.  Hence this term is exponentially suppressed at large $\ell$ %and we can ignore this term. Hence we 
    and we are left with,
   \begin{align}\label{ghat}
    {\mathcal{G}}_{\Delta, \ell}(u, v) & \sim \frac{1}{2^{\ell}}\, u\, k_{\DD+\ell}(\bz)\,k_{\DD-\ell-2}(z),
   \end{align}
   %We see that,
  where,
   \be
   k_{\DD-\ell-2}(z) =u^{\frac{\tau}{2}} +\cdots
   \ee
  and the dots indicate higher order terms in $u$. For $k_{\DD+\ell}(\bz)$ let us consider the integral representation of the hypergeometric function in the limit $\ell \rightarrow \infty$,
  \begin{align}
   _2F_1  \bigg[\frac{\DD+\ell}{2}, \frac{\DD+\ell}{2}, \DD+\ell, \zb \bigg] &= \,  _2F_1  \bigg[\frac{\tau}{2}+\ell, \frac{\tau}{2}+\ell, \tau+2\ell, 1-v \bigg] \nn & 
   = \frac{\Gamma(2\ell)}{\Gamma^2(\ell)} \int_{0}^{1} \frac{dt}{t(1-t)}\,\bigg(\frac{t(1-t)}{1-t\,v}\bigg)^{\ell}\nn
   & \approx \frac{2^{2 \ell-1} \sqrt{\ell}}{\sqrt{\pi }} \, \int_{0}^{1} dt\, \frac{t^{\ell-1}}{1-t}\, e^{-\frac{t\,y}{(1-t)\,y}}\,.
  \end{align}
  Now we define a new variable $s=\frac{t}{1-t}$ and rewite the integral as,
  \begin{align}
  _2F_1  \bigg[\frac{\tau}{2}+\ell, \frac{\tau}{2}+\ell, \tau+2\ell, 1-v \bigg]  & \sim \frac{2^{2 \ell-1} \sqrt{\ell}}{\sqrt{\pi }}  \int_{0}^{\infty} \frac{ds}{s}\, e^{-\frac{s\, y}{\ell}-\frac{\ell}{s}} \nn &=  \frac{2^{2 \ell} \sqrt{\ell}}{\sqrt{\pi }} K_0 (2\ell \sqrt{v}) +O(1/\ell)\,.
  \end{align}
   Now we will see how the derivative of \eqref{ghat} behaves in the large $\ell$ limit.
   \begin{align}\label{g2}
 \partial_{\DD}  {\mathcal{G}}_{\Delta, \ell}(u, v) & \sim \frac{1}{2^{\ell}}\, u\, \bigg[\half\, \log \zb \, k_{\DD+\ell}(\bz)\,k_{\DD-\ell-2}(z)+\half\, \log z \, k_{\DD+\ell}(\bz)\,k_{\DD-\ell-2}(z)\nn&+
  \zb^{\frac{\DD+\ell}{2}}\, \partial_{\DD}\,_2F_1  \bigg[\frac{\DD+\ell}{2}, \frac{\DD+\ell}{2}, \DD+\ell, \zb \bigg]\,k_{\DD-\ell-2}(z) \nn & +z^{\frac{\DD-\ell-2}{2}}k_{\DD+\ell}(\bz)\, \partial_{\DD}  \,_2F_1\bigg(\frac{\DD-\ell-2}{2}, \frac{\DD-\ell-2}{2}, \DD-\ell-2, z\bigg) \bigg]\,.
   \end{align}
  Note that,
  \be
  \log \zb =\log (1-v)= -v+O(v^2).
  \ee
  Hence the term involving $\log \zb$ in \eqref{g2} is subleading in $v$ and can be ignored. Now we look at the derivatives of the $_2F_1$.
  \begin{align}
&\partial_{\DD}  \,_2F_1\bigg(\frac{\DD-\ell-2}{2}, \frac{\DD-\ell-2}{2}, \DD-\ell-2, z\bigg) \nn&= \sum_{n=0}^{\infty} \frac{z^n\, \Gamma (\tau -2)  \Gamma \left(n+\frac{\tau }{2}-1\right)^2}{n! \,\Gamma \left(\frac{\tau -2}{2}\right)^2 \Gamma (n+\tau -2)}\left(H_{n+\frac{\tau }{2}-2}-H_{n+\tau -3}-H_{\frac{\tau }{2}-2}+H_{\tau -3}\right).
  \end{align}
  The leading term vanishes for $n=0$ and this is again subleading in $u$. Hence, we can ignore this term as well. We will finally focus on the second line of \eqref{g2}. Using the integral representation of the hypergeometric function we get,
  \begin{align}
\partial_{\DD}\,_2F_1  \bigg[\frac{\DD+\ell}{2}, \frac{\DD+\ell}{2}, \DD+\ell, \zb \bigg]&= \partial_{\DD} \int_{0}^{1} dt\,  \frac{\Gamma (\ell+\Delta ) (-(t-1) t)^{\frac{1}{2} (\Delta +\ell-2)} (1-t {\zb})^{\frac{1}{2} (-\Delta -\ell)}}{\Gamma \left(\frac{\ell+\Delta }{2}\right)^2} \nn
&= \int_{0}^{1} dt\, \frac{(-(t-1) t)^{\ell+\frac{\tau }{2}-1} \Gamma (2 \ell+\tau ) \log \left(\frac{(t-1) t}{t {\zb}-1}\right) (1-t {\zb})^{-\ell-\frac{\tau }{2}}}{2 \Gamma \left(\ell+\frac{\tau }{2}\right)^2}\nn
& - \int_{0}^{1} dt\, \frac{(-(t-1) t)^{\ell+\frac{\tau }{2}-1} \Gamma (2 \ell+\tau ) (1-t {\zb})^{-\ell-\frac{\tau }{2}}}{\Gamma \left(\ell+\frac{\tau }{2}\right)^2} \nn& \times  \left(\psi\left(\ell+\frac{\tau }{2}\right)-\psi (2 \ell+\tau )\right).
  \end{align}
  In order to do the integral we define a new variable $s= \frac{t}{1-t}$ and rewrite the first integral as follows,%{\bf PD: Explain why both the integrals give the same result!}
  \begin{align}
  &\int_0^{\infty } {s\, y \,e^{-\frac{s\,y}{\ell}-\frac{\ell}{s}}}\bigg(\frac{1}{\ell^2 s}+\frac{1}{s^2}\bigg) \, ds 
  ={4 \sqrt{v} \,K_1\left(2 \sqrt{y} \right)}+{\rm{subleading \, terms}}\,.
  \end{align}
 % {\bf PD: Comment on the dotted terms}
  Hence this is suppressed in $v$ and  can be ignored.
  In the $\ell \rightarrow \infty $ limit,
  \be
  2 \psi (2 \ell+\tau )-2 \psi \left(\frac{1}{2} (2 \ell+\tau )\right) \sim 2\Log 2+ O(1/\ell)\,.
  \ee
  Hence, we have
  \be
 \sum_{\ell \gg 1} 2\,q_{_0, _\ell} \,\partial_{\DD}{{G}}_{\Delta, \ell}(u, v) =  2 \bigg(\frac{u}{v^2}\bigg)^{\frac{\DD_\f}{3}}\, \log 2\,.
  \ee
 %  \section{Useful integral}\label{int}
  %\be\label{int1}
%  \int_{\ell_0}^{\infty} \ell^{2\DD_\f-1}\, K_0(2\ell\,\sqrt{v}) = \frac{v^{-\DD_\f}}{4\,\Gamma^2(\DD_\f)} +\cdots
%  \ee
%  \be\label{int2}
%  \int_{\ell_0}^{\infty} \ell^{2\DD_\f-1}\, K_0(2\ell\,\sqrt{v})\,\log \ell =- \frac{v^{-\DD_\f}}{8\,\Gamma^2(\DD_\f)} \bigg(\log v-2 \psi (\Delta_\phi )\bigg)+\cdots
%  \ee 
 % \begin{align}\label{int3}
% & \int_{\ell_0}^{\infty} \ell^{2\DD_\f-1-a}\, K_0(2\ell\,\sqrt{v}) = \frac{v^{-\DD_\f+\frac{a}{2}}\,\Gamma^2(\DD_\f-a/2)}{4} +\cdots\nn
  %\ee 
  %\be\label{int4}
 %& \int_{\ell_0}^{\infty} \ell^{2\DD_\f-1-a}\, K_0(2\ell\,\sqrt{v})\,\log \ell =- \frac{v^{-\DD_\f+a/2}\,\Gamma^2(\DD_\f-a/2)}{8} \bigg(\log v-2\, \psi (\Delta_\phi-a/2 )\bigg)+\cdots\,.
  %\end{align}
 % where the $\cdots$ are subleading in $v$\,.
  \section{\label{sec:int}Some details of the bulk computation}
  Let's start with the solution \eqref{eq:ansatz} for the differential equation \eqref{eq:inhomDE}, 
  \begin{align} \label{eq:soln}
  \Phi(x) &= \int \sqrt{-g} \, d^{d+1}x' \, G(x, x') \, \Phi_{\text{CFT}}(x') \nn 
  &= \int \, \sqrt{-g}  \, dt' d{\rho}' d{\Omega}' \, \int \frac{d \omega}{2 \pi i} \sum_{n, \ell, \vec{m}}  \, \frac{e^{i \omega (t - t')}}{\omega^2_{n, \ell} - \omega^2} \, \phi^*_{n, \ell, \vec{m}}(\vec{x}) \, \phi_{n, \ell, \vec{m}}(\vec{x}')\nn
  & \times  \quad \frac{1}{N_{\Delta, n, \ell}} \, e^{i \omega_{n, \ell} t'} \, Y_{\ell J}(\Omega') \sin^{\ell}{\rho'} \cos^\Delta{\rho'} \,  _2F_1 \bigg(-n, \Delta + n + \ell, \ell+ \frac{d}{2}, \sin^2{\rho'}\bigg) \,.
  \end{align} 
  Although we write \eqref{eq:soln} for arbitrary $d$, we will work in $d=4$ in what follows. The above integration is over all coordinates \emph{i.e.} $t', \rho'$ and $\Omega'$.  Performing the $t'$ integral first and then the $\omega$ integral we get,
  \begin{align} 
  \Phi(x) &= \int \, \sqrt{-g} \, d{\rho}' d{\Omega}' \, \sum_{n', \ell', \vec{m'}}  \, \frac{\phi^*_{n', \ell', \vec{m}'}(\vec{x}) \, \phi_{n', \ell', \vec{m}'}(\vec{x}')}{i (\omega^2_{n', \ell'} - \omega^2_{n,\ell})} \, \nn
  & \times \quad \frac{ e^{i \omega_{n, \ell} t}}{N_{\Delta, n, \ell}} \, \, Y_{\ell J}(\Omega') \sin^\ell {\rho'} \cos^\Delta{\rho'} \,  _2F_1 \bigg(-n, \Delta + n + \ell, \ell+ \frac{d}{2}, \sin^2{\rho'}\bigg) \,.
  \end{align} 
  
  The modes $\phi_{n, \ell, \vec{m}}(x)$ are known functions,
  \begin{align} 
  \phi_{n, \ell, \vec{m}}(x) = \, Y_{\ell m}(\Omega) \sin^\ell {\rho} \cos^\Delta{\rho} \,  P_n^{\ell + d/2 -1, \nu}(\cos{2 \rho}),
  \end{align}
  where $P_n^{m, \nu}(\cos{2 \rho})$ are Jacobi polynomials. Next we use the orthogonality of spherical harmonics (see \emph{e.g.} complement $A_{\text{VI}}$ of \cite{Cohen-Tannoudji:101367})
  \begin{align} 
  \int  d \Omega' Y^*_{\ell' m'}(\Omega') Y_{\ell J}(\Omega')  = \delta_{\ell, \ell'} \delta_{J m'},
  \end{align}
  to obtain,
  \begin{align} 
  \Phi(x) &= \int \, \sqrt{-g} \, d{\rho}' \, \sum_{n'}  \, \frac{1}{(\omega^2_{n', \ell} - \omega^2_{n,\ell})} \,
  P_{n'}^{\ell + d/2 -1, \nu}(\cos{2 \rho}) P_{n'}^{\ell + d/2 -1, \nu}(\cos{2 \rho'}) \nn
  &\times Y^*_{\ell J}(\Omega)  (\sin{\rho})^\ell (\cos{\rho})^\Delta \frac{ e^{i \omega_{n, \ell} t}}{i N_{\Delta, n, \ell}}  (\sin{\rho'})^{2 \ell} (\cos{\rho'})^{2 \Delta}  \,  _2F_1 \bigg(-n, \Delta + n +\ell, \ell+ \frac{d}{2}, \sin^2{\rho'}\bigg) \nn
 &= \frac{Y^*_{\ell J}(\Omega) e^{i \omega_{n, \ell} t}}{i N_{\Delta, n, \ell}} \frac{n!}{(\ell +\frac{d}{2})_n} (\sin{\rho})^\ell (\cos{\rho})^\Delta \sum_{n'} \frac{1}{(\omega^2_{n', \ell} - \omega^2_{n, \ell})}P_{n'}^{\ell + d/2 -1, \nu}(\cos{2 \rho})  \nn
& \times \int_0^{\pi/2} d{\rho}'   (\sin{\rho'})^{2 \ell-d-1} (\cos{\rho'})^{2 \Delta+d-1}  P_{n'}^{\ell + d/2 -1, \nu}(\cos{2 \rho'}) P_{n}^{\ell + d/2 -1, \nu}(\cos{2 \rho'}) \nn \label{eq:ansatz1}
%  &= \frac{Y^*_{lJ}(\Omega) e^{i E_{n, l} t}}{i N_{\Delta, n, l}} \frac{n!}{(l+\frac{d}{2})_n} (\sin{\rho})^l (\cos{\rho})^\Delta \sum_{m} \frac{1}{(E^2_{m, l} - E^2_{n,l})}P_{m}^{l + d/2 -1, \nu}(\cos{2 \rho})  \nn
%  & \quad \underbrace{\int_0^{\pi/2} d{\rho}'   (\sin{\rho'})^{2 \alpha +1} (\cos{\rho'})^{2 \beta -1}  P_{m}^{\alpha, \nu}(\cos{2 \rho'}) P_{n}^{\alpha, \nu}(\cos{2 \rho'})}_{I^{\rho'}}
  &\equiv  \frac{Y^*_{\ell J}(\Omega) e^{i \omega_{n, \ell} t}}{i N_{\Delta, n, \ell}} \frac{n!}{(\ell + \frac{d}{2})_n} (\sin{\rho})^\ell (\cos{\rho})^\Delta \, f(\rho)\,,
  \end{align} 
  where $\sqrt{-g}  = (\sin{\rho'})^{-d-1}\, (\cos{\rho'})^{d-1}$, and $\alpha = \ell + \frac{d}{2} -1; \quad \beta = \Delta - \frac{d}{2} = \nu $. 
  
To obtain the solution $\Phi(x)$ one needs to compute $f(\rho)$.   One way would be to perform the $n'$ sum and $\rho'$ integral. Here we take an alternative approach, namely we substitute \eqref{eq:ansatz1} to \eqref{eq:inhomDE} as an ansatz to get the following  differential equation for $f(\rho)$, 
\begin{align} \label{eq:f-ODE}
 f''(\rho ) - \big( (2 \Delta -3) \tan \rho - (2 \ell  + 3) \cot \rho \big) f'(\rho ) - (\Delta -\omega +\ell ) (\Delta +\omega +\ell ) f(\rho )  =\sec ^2 \rho.
\end{align}
  
  All one needs to do is to solve for $f(\rho )$. The equation \eqref{eq:f-ODE} is an second-order inhomogeneous ODE. Let's make the following change of variables, 
  $$ z = \cos^2 \rho. $$
  Above equation \eqref{eq:f-ODE} reduces to, 
    \begin{align} \label{eq:f-ODE-z}
(1-z) z \,  f''(z) +\big(c -(a + b +1) z \big) \,  f'(z)  - a b f(z) = z^{q},
  \end{align}
  with,
   \begin{align}
a &= \frac{1}{2} (\Delta -\omega +\ell ) = -n, \nn 
b &= \frac{1}{2} (\Delta +\omega +\ell ) = n +\ell +  \Delta \nn 
c &= \Delta -1 \nn
q &= -1.
  \end{align}
  The corresponding homogeneous ODE is the standard hypergeometric differential equation,
  \begin{align}
(1-z) z \,  f''(z) +\big(c -(a + b +1) z \big) \,  f'(z)  - a b f(z) = 0,
  \end{align}
with the following solution, %(\textcolor{blue}{assuming $1-c = 2 - \Delta$ is not an integer}),
\begin{align} \label{eq:fsol-homo}
f^h (z) &= \, C_1 \ _2F_1 (a, b, c; z)  +  C_2 \ z^{1-c} \, _2F_1 (b - c + 1, a - c + 1, 2- c; z)\,.
\end{align}
Imposing regularity at the centre of AdS ($z \to \infty$), forces one to choose $C_2 =0$. Other boundary condition fixes the normalization $C_1 = \mathcal{N}_{\Delta, \ell, m}$. The inhomogeneous DE \eqref{eq:f-ODE}, has the following particular solution\footnote{See eqn (12) and eqn (13) of  \cite{Ancarani:2009zz}.},
 \begin{align} \label{eq:fsol-inhomo}
f^p (z) &=  \frac{\Gamma(1 + q) \, \Gamma (c-1)}{\Gamma (c)} \, _2F_1(a,b;c;z)\,.
 \end{align}
 Notice that the solution \eqref{eq:fsol-inhomo} is well behaved for all values of $q$, except for $q \in \mathbb{Z}^{^-}$.  One can analytically continue to complex $\mathbf{q} = q_1 + i q_2$. Then for any negative integer $\mathbf{q} =  - p$, one can expand $\Gamma(1 + \mathbf{q})$ in small complex neighborhood as following\footnote{Near any simple pole at $z = - n$, where $n \in \mathbb{Z}^{^+}$, $$\Gamma(- n + z) = \frac{(-1)^n}{n!} \bigg( \frac{1}{z} + \psi(n +1) + \mathcal{O}(z) \bigg). $$},
  \begin{align} 
  \Gamma(1 + \mathbf{q}) &=  \Gamma(1 - p +  i \epsilon) \nn
  &= \frac{(-1)^{p-1}}{(p-1)!} \, \left(\frac{1}{i \epsilon } + \psi (p) + \mathcal{O}(\epsilon) \right)\,.
  \end{align}
 It is evident that in the limit $\epsilon \to 0$, only the imaginary part blows up, whereas the real part is independent of $\epsilon$\footnote{The function $\Gamma(1 + \mathbf{q})$ is analytic on the complex $\mathbf{q}$-plane with $q \in \mathbb{Z}^{^-}$ removed -- which is an open set.  Therefore one can approach the disconnected singular points at $q \in \mathbb{Z}^{^-}$  from any directions in the complex $\mathbf{q}$-plane. The finite part will be independent of the cut-off.}. Thus there is a consistent prescription of extracting $\epsilon$ independent value as follows 
   \begin{align} 
   \Gamma(1 - p) = \frac{(-1)^{p-1}}{(p-1)!} \, \psi (p).
   \end{align}
 Here we are particularly interested in $q = -1$ \emph{i.e.} $p=1$, for which 
    \begin{align} 
    \Gamma(1 + q) =  \psi (1)  =  - \,  \gamma_{_E}.
    \end{align}
  Therefore the particular solution reduces to, 
  \begin{align} \label{eq:fsol-inhomo-reg}
 f^p (z) &= - \,  \gamma_{_E} \,  \frac{\Gamma (c-1)}{\Gamma (c)} \, _2F_1(a,b;c;z).
 \end{align}

 The argument of the hypergeometric function is $z = \cos ^2 \rho$. Since we want the solution with the variable  $\sin ^2 \rho$, let's use the following identity,
   \begin{align} 
  _2F_1(-m ,b ; c ; z) = \frac{(c-b)_m}{(c)_m} \, _2F_1(-m ,b ; b-c-m+1 ; 1- z).
  \end{align}
 to get,
    \begin{align} \label{eq:fsol-inhomo-reg-sine}
  f^p (z) &= - \, \gamma_{_E} \, \frac{\Gamma (c-1)}{\Gamma (c)} \, \, \frac{(c-b)_n}{(c)_n} \, _2F_1(-n, b ; b - c -n +1  ; 1- z) \nn 
 &=  - \, \gamma_{_E} \, \frac{\Gamma (\Delta -2) \, \Gamma (n+ \ell + 2)}{\Gamma (n+\Delta -1) \,  \Gamma (\ell +2 )} \, _2F_1\left(-n,n+\ell +\Delta ;\ell +2;\sin ^2 \rho  \right)\,.
  \end{align}
 
  Now we have all the ingredients to write down the full solution to  \eqref{eq:inhomDE}, 
 \begin{align} 
 \Phi(x) &=   - \, \gamma_{_E} \,  \frac{Y^*_{lJ}(\Omega) e^{i \omega_{n, \ell} t}}{i N_{\Delta, n, l}}  (\sin{\rho})^\ell (\cos{\rho})^\Delta \frac{n!}{(\ell+\frac{d}{2})_n}  \, \frac{\Gamma (\Delta -2) \, \Gamma (n+ \ell +2)}{\Gamma (n+\Delta -1) \,  \Gamma (\ell +2)} \nn & \quad \quad \times \, _2F_1\left(-n,n+\ell +\Delta ;\ell +2;\sin ^2 \rho  \right)\,.
 \end{align}

\bibliographystyle{bibstyle}
\bibliography{LogCFTs}
\end{document}